# Title: Temporal interleaving artifacts in spiral MRSI: characterization and retrospective correction

Antoine Naëgel[1,2], Magalie Viallon[1,3], Thomas Troalen[2], Christopher T. Rodgers[4], Dinesh K Deelchand[5], Paul Nobre[1], David Porter[6], Jabrane Karkouri[2,4], Hélène Ratiney[1]

[1] INSA-Lyon, Université Lyon 1, UJM-Saint Etienne, CNRS, Inserm, CREATIS UMR 5220, U1294, F-69621, LYON, France

[2] Siemens Healthcare SAS, Courbevoie, France

[3] Radiology department, UJM-Saint-Etienne, Centre Hospitalier Universitaire de Saint-Étienne, Saint-Etienne, France

[4] Wolfson Brain Imaging Centre, Department of Clinical Neurosciences, University of Cambridge

[5] Center for Magnetic Resonance Research, Department of Radiology, University of Minnesota, Minneapolis, USA

[6] Imaging Center of Excellence, University of Glasgow, Glasgow, UK

Corresponding author: Hélène Ratiney, helene.ratiney@creatis.insa-lyon.fr

List of abbreviations:

1. **ADC** : Analog-to-Digital Converter
2. **CPP** : Comité de Protection des Personnes (Institutional Review Board )
3. **NCT** : National Clinical Trial
4. **SD** : Standard Deviation
5. **LCModel** : Linear Combination of Model spectra
6. **IMCL**: intra-myocellular lipids

## Abstract:

**Purpose:** In Spiral Magnetic Resonance Spectroscopic Imaging (MRSI), achieving suitable spatio-temporal resolutions requires interleaving. Sampling inconsistencies occurring between temporally interleaved signals, —whether due to time-interleaved ADCs or phase/frequency mismatches in the excitation and demodulation stages—**,** result in artifacts in the recombined data. This paper proposes a mathematical description of these unavoidable artifacts and a post-processing solution to circumvent the issues and mitigate them.

**Methods:** The study proposed a detailed description and explanation of these artifacts using data acquired with MRSI interleaved sequence with gradients off, as well as standard interleaved FID sequence. The random signal mismatch between the interleaves was described as additional artifactual signals $\psi(t)$ at each temporal interleave. A model was further proposed to match the acquired signal in relation to the ideal, artifact-free signal. Considering M interleaves, a correction method was derived relying on estimating 4 parameters in the $\psi(t)$ model for each interleave by minimizing a multivariable cost function. This correction method was evaluated on spectra acquired from phantoms and in vivo acquisitions obtained in healthy volunteers.

**Results:** The proposed correction method significantly reduced artifacts by a factor of 4 in the phantom and 1.8 in the healthy volunteers. This correction ensured accurate spectral quantification in regions where artifacts overlapped with the content of interest, such as lipids in subcutaneous fat. Additionally, it was demonstrated that appropriately selecting the number of temporal interleaves can shift the artifact away from the frequency of interest, leveraging the periodic and limited frequency span of the observed artifacts.

**Conclusion:** In spiral-MRSI, temporal interleaves, while enhancing spectral bandwidth, introduced spurious content characterized by distinctive resonance frequencies and non-reproducible amplitudes and phases. The study proposes two solutions: manipulating the number of interleaves to control artifact frequency localization or using a retrospective correction method to approximate and attenuate artifacts.

## Introduction

Magnetic resonance spectroscopic imaging (MRSI) is a powerful tool for quantitatively exploring living organisms as it provides spectroscopic information in addition to the image generated by MRI alone[1–6]. However, its use in clinical practice remains limited due to long acquisition time and complex data processing. The recent development of spatio-temporal encoding techniques helped to shorten the acquisition time to ease the use of MRSI in clinical applications[7–11].

In the case of spiral MRSI, spatial and temporal interleaving is used to achieve suitable spatiotemporal resolutions[12,13]. Indeed, an acquisition utilizing a spiral encoding technique has a minimum temporal sampling step given by one spiral length, which limits the resulting spectral bandwidth. Temporal interleaving is performed to increase spectral bandwidth and consists of repeating the spiral train acquisition shifted by a time equivalent to the desired temporal resolution (Figure 1). Once the data have been acquired, all the different acquisition sets are combined to reconstruct the data at the targeted temporal resolution.

Signal instabilities occurring between interleaved signals are breaking the mandatory signal coherence across interleaves for cleaned temporal recombination, further translating into peak artifacts in the reconstructed spectra. Such artifacts and their consequences are very well described and addressed in the field of electronic circuits and systems[14–16]. However, to the best of our knowledge, these artifacts have only been mentioned in the literature, for instance by Bogner et al[6] and Hingerl et al. [17], particularly in the context of ultra-high-field magnetic resonance when attempting to increase spectral bandwidth using temporal interleaving without water suppression. Moreover few publications reported temporal interleaving but with small interleaves (~2), large sampling steps, and relatively low spatio-temporal bandwidth[3,11,18–22]. When higher spatial resolution is desired, the spiral readout duration increases, decreasing spectral resolution. A higher order of temporal interleaving is thus necessary to maintain high spectral resolution, making the technique extremely sensitive to signal incoherence across shots.

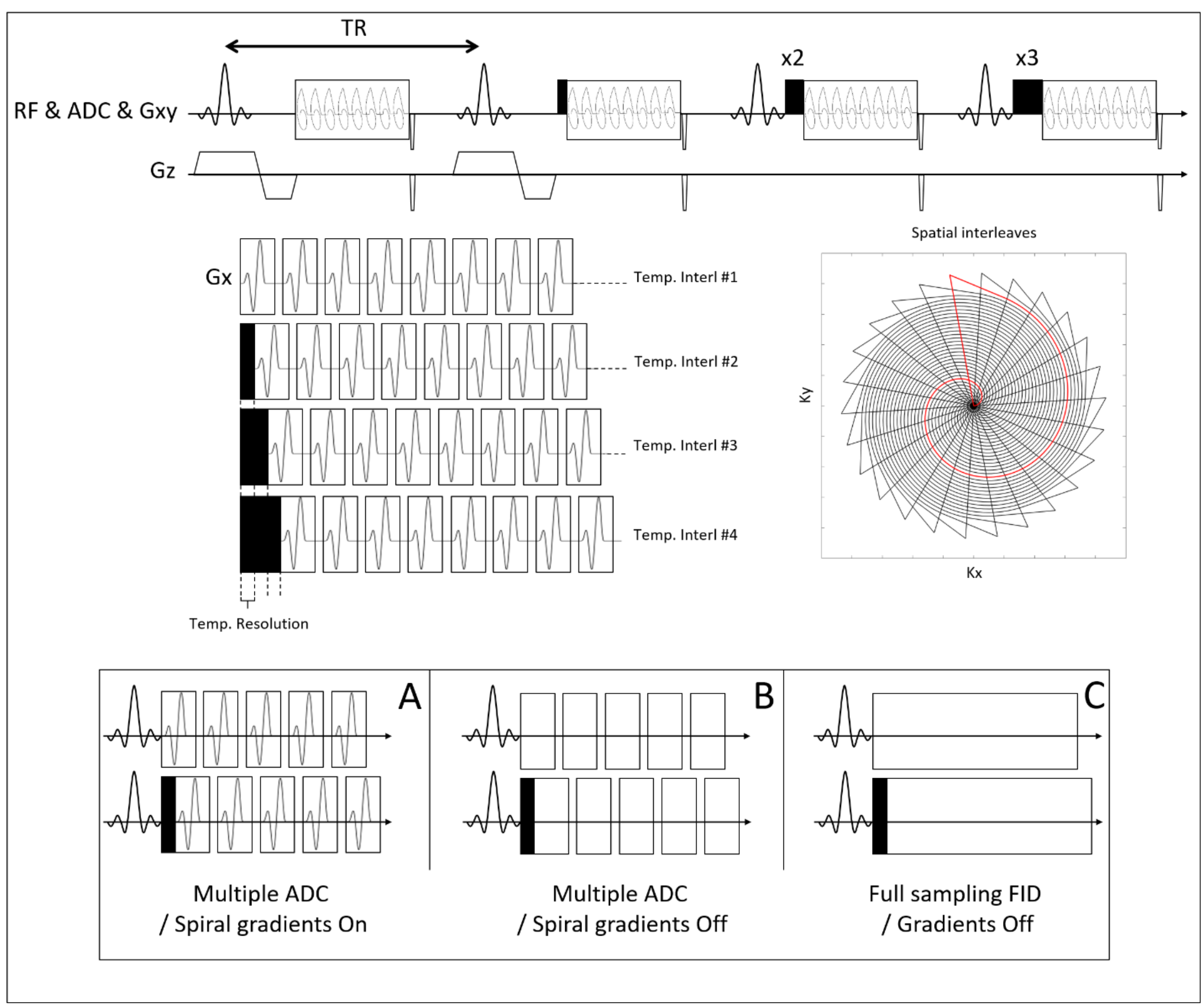

***Figure 1: A constant-density spiral readout is used with temporal interleaving in the imaging plane.*** *After a slice-selective excitation pulse, spiral gradient modulations follow. These spiral trajectories are repeated to cover the entire free induction decay in time. Due to the long duration of a spiral, temporal interleaves (illustrated with four here) are acquired. Each temporal interleave is obtained after a separate excitation, and identical spiral gradients are played out with a predefined delay. This delay sets the dwell time, determining the spectral readout bandwidth (2k Hz). Spatial interleaving is employed to encompass the entire k-space. A-B-C is an overview of the different sequences implemented to explore artifacts: (A) the conventional spiral MRSI sequence with multiple ADCs and gradients on, (B) the same sequence but with gradients off, and (C) a conventional FID spectroscopy sequence with a time shift to create temporal interleaving.*

This paper aims to review, extensively describe and illustrate the signal incoherence and the resulting artifacts that arise when employing temporal interleaving in spiral MRSI sequences, with ADC time-interleaving and repeated opening/closing of ADCs. It outlines the methods we deployed for mathematically characterizing these artifacts and offers a post-processing solution to compensate for them. Finally, we demonstrate, using illustrative in vitro and in vivo examples, the efficiency of the correction of this artifact through the proposed post-processing procedure.

## Material and Method:

A FID-MRSI Spiral prototype sequence was developed on a 3T MR scanner (MAGNETOM PRISMA, software version N4_VE11C, Siemens Healthineers, Germany) using a spatially selective RF pulse followed by a readout gradient consisting of a spiral train for spiral filling of the k-t space. It was tested sequentially without water suppression and in different modes (gradients on/off and multiple/single ADC) or parameter settings (number of temporal/spatial interleaves) as described in Figure 1. In addition, a standard (none spatially resolved) FID sequence was used to acquire a simple time signal with full sampling and to build interleaved signals and investigate the artifact behaviour in a multi-ADC like simple FID configuration (Supplementary material Figure S1).

These different modes (gradients on/off and multiple/single ADC) were acquired on a spectroscopy phantoms (spherical “Braino” phantom (GE Medical Systems, Milwaukee, WI, USA) ) and healthy volunteers, who provided written informed consent and were recruited under a protocol approved by the institutional review boards (CPP Nord Ouest VI, # 19.02.22.52507, and CPP Ile de France VIII, #20 04 05) and registered at ClinicalTrials.gov (NCT04363606, NCT05031598). Acquisition parameters are summarized in Table S1 in the supplementary material. All reconstructions of the scanner raw data were performed offline in Matlab (The Mathworks, version R2021a).

### *Origin and importance of the artifact and proposed mathematical model*

Figure 2A shows a spectrum acquired on the phantom and typically affected by artifacts after recombining several temporal interleaves. The recombined time domain signal represented in Figure 2B shows the signal incoherencies at the origin of the artifactual peaks observed in the spectrum (i.e., repeated signal jumps, as a function of the number of temporal interleaves) while simple signal decay attributed to water signal is expected. Figure 2C showcase a spectrum acquired in vivo to appreciate how these artifacts practically impact the observed signal (here the lipid region) of interest and understand the importance of it.

Acquisitions were repeated several times to study the coherence, repeatability, and interleaving across scans. Finally, the sequence was tested on two different 3T scanners running the same software version (VE11C) to rule out site-specific hardware problems.

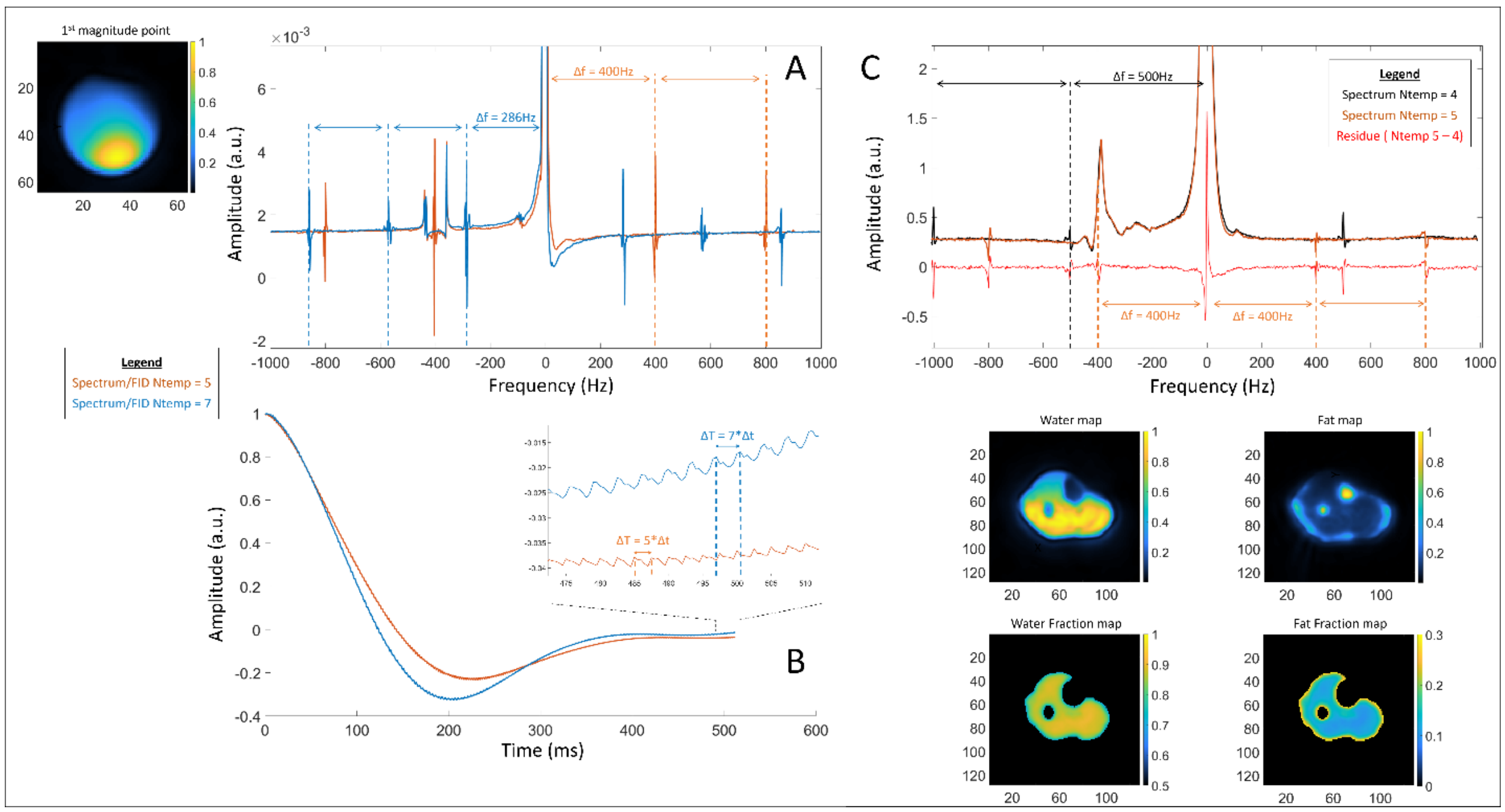

***Figure 2: Artifacts in spectroscopic (A-C) and temporal (B) dimension with different number of temporal interleaves to see the variation of the drift in frequency and signal jumps in FID.*** *The acquisitions were carried out on the "Braino" phantom, and in vivo on the lower part of the leg of a healthy volunteer (supplementary material table). In the frequency dimension, the frequency delta between artifact peaks changes, and can be characterized as follows: Δf = 1/(M x Δt).In the time dimension, the temporal delta between signal jumps can be characterized as follows : ΔT = M x Δt. Spectra acquired with 5 temporal interleaves present artifacts in a region of interest (lipids). NTemp = Number of Temporal Interleaves, and 4 dummy scans.*

Using acquisitions with the gradients deactivated (Figures 1B-C), one can extract and examine the behavior of an additional artifactual signal for each interleave, which we will call $\psi_m(t)$ in the following for each $m$ interleave. This characterization procedure is described in Figure 3. The sampled points in the first interleave correspond to the time points that would have been sampled by a spiral train. Thus, across different sections with gradients turned off, the first interleave contains all the points necessary to reconstruct a spectrum (the black spectrum in Figure 3) with a 2 kHz bandwidth, corresponding to a 500 µs sampling interval. This spectrum can be compared to the one obtained by combining multiple interleaves, as would occur during the temporal interleaving of a spiral train with gradients on. By subtracting one from the other, a signal representative solely of the artifact can be extracted and analyzed in the time domain for each interleave (Figure 3).

To further explain and describe in detail the origin and effects of the artifacts discussed in this paper, we conducted additional experiments based on simple FID acquisitions, ensuring that any issues related to the implementation of advanced sequences, such as spiral sequences, could be ruled out. All these sequence variants are described and analyzed in Figures S1 and S2 of the supplementary material. Indeed conventional FID acquisitions can be temporally interleaved to create datasets that reproduce artifacts (Figure S1, left column). We also performed artificial temporal interleaving of the signal based on accumulated acquisitions, intrinsically acquired without temporal shift in the acquisition, as the ADC starts at the same time (Figure S1, right column). This last experience aims at illustrating the random nature of the artifacts (figure S2).

These sequences (spiral MRSI gradient off or FID sequences) are only used to explore the behavior of the artifact and are not intended to be a pre-scan correction technique since artifact resulting from multiple temporal interleaving is not reproducible and cannot be pre-calibrated for each system. Therefore, we propose a correction post-acquisition. We use a model $\widehat{s_{acq}}$ to describe the signal acquired with respect to the ideal signal (without artifacts) *s*, whose component with the greatest amplitude is centered at 0Hz. For *M* interleaves, a final dwell time $\Delta t$ (bandwidth of the resulting spectrum being $\frac{1}{\Delta t}$), the sampling step of each interleave (related in case of spiral MRSI to the spiral length) $\Delta T$ (with $\Delta T = M\Delta t$), the time domain acquired signal can be written as follows:

$$\widehat{s_{acq}}(nt) = \sum_{m=0}^{M-1} \psi_m\,(k\,\Delta T) + s(k\Delta T + m\Delta t\,),\ with\ n=kM+m \text{ and } 0 \leq n \leq N \qquad [1]$$

With *n* being the index of the signal sample acquired with a step size of $\Delta t$ , and *N* the total number of temporal samples (taking into account all interleaving). For $\psi_m$, we propose as a first approximation and from observations (Figure 3D):

$$\psi_m(t) = A_m \exp(-\alpha_m \mathrm{t}) \exp(i2\pi f_m t + i\phi_m) \qquad [2]$$

With the following 4 parameters $A_m$, $\alpha_m$, $f_m$, et $\varphi_m$ defining signal amplitude, damping, frequency, and phase respectively, leading to 4M parameters when considering M interleaves.

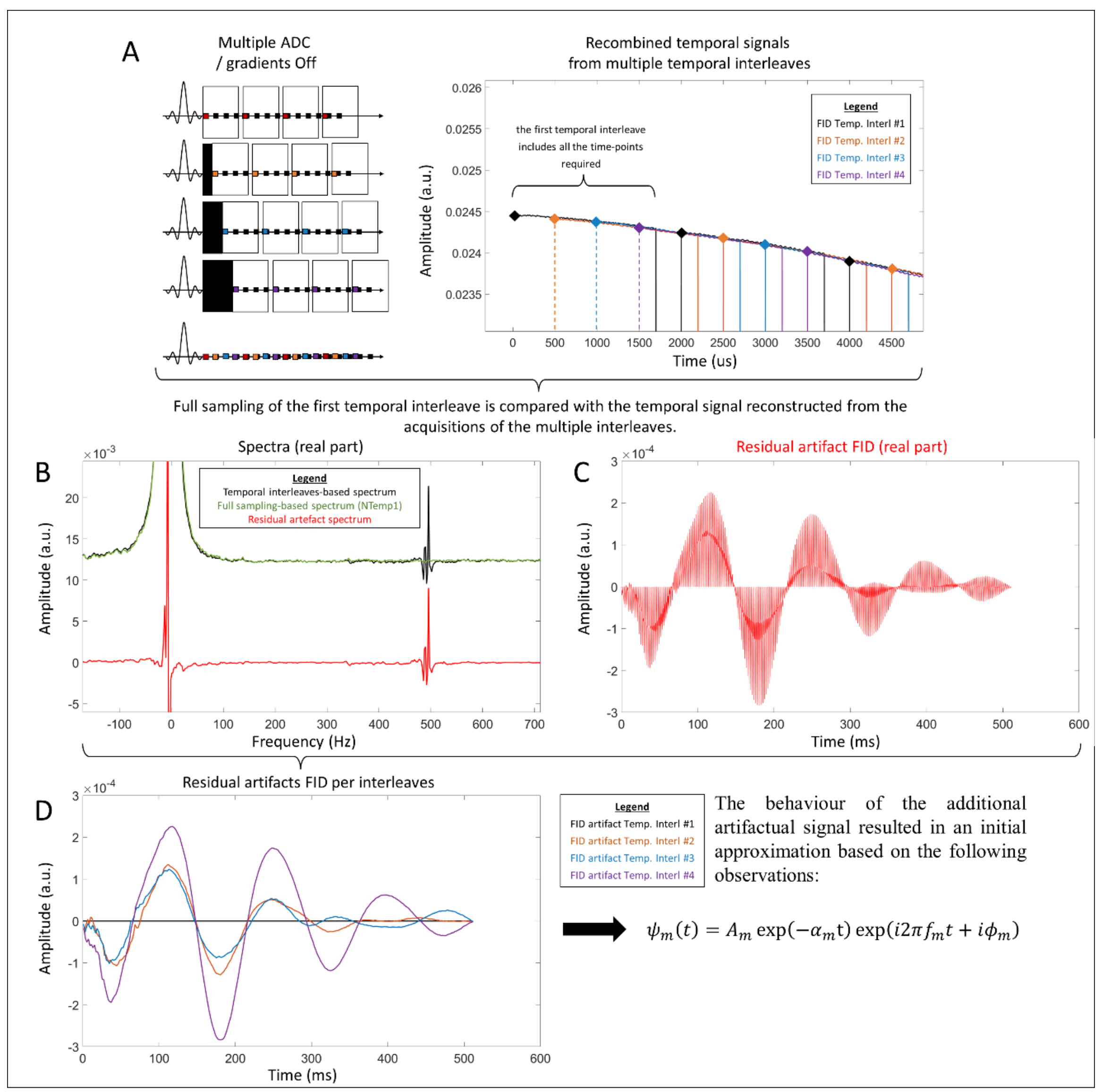

***Figure 3: Method for characterizing artifacts from gradient-off acquisitions in a spiral sequence.*** *Gradient-off acquisitions presented in (A-left) offer the advantage of coherent signal sections between temporal interleaves, as shown in the plots in magnitude in (A-right). In (B) a spectrum (green) obtained exclusively from the first interleave is displayed, and thus resulting from a single ADC opening. In this case, all sampled time points at 500 µs are acquired within the first interleave (in a spiral sequence, with gradient-on these points are encoded also in the k-space). Additionally, another spectrum (black), generated through time interleaving (with a separate ADC opening for each interleave), is shown. This spectrum contains the artifact under investigation. By subtracting the green spectrum (obtained with a single ADC opening) from the black spectrum, a proxy for the artifact (red) can be extracted. This artifact, represented in the frequency domain, is also observable in the time domain after an inverse Fourier transform, as shown in (C). Finally, decomposing this artifact by temporal interleaving reveals its characteristic time domain signature in (D). Note that in this demonstration the first interleave of the artifact does not contain signal as it is used as a reference.*

Note that $\psi_m$ is independent from the delay $m\Delta t$. With the arbitrary occurrence of $\psi_m$ at each interleave, artifactual signal variations may occur at sampling $\Delta t$, $2\Delta t$ ... $M\Delta t$ ; resulting in peak contributions at frequencies +/- $\frac{1}{\Delta t}, \frac{1}{2\Delta t} \cdots \frac{1}{M\Delta t}$.

It is also important to highlight that it is well established that, the initial data points collected immediately after the opening of the ADC show a distinctive behavior characterized by signal jumps. In sequences involving encoding gradients, such spiral gradients, it is necessary to introduce a delay in the application of these gradients by a few microseconds after the ADC opening[23]. This delay ensures that the signal is acquired in a more stable region. While these anomalies are widely recognized within the community, they are infrequently discussed in the imaging literature. If not properly addressed, these mismatches in the signal can lead to more severe artifacts. In this study, and specifically for Figure 3, this delay has been implemented to ensure the most optimal version of the sequence.

*Retrospective correction of the artifacts*

The proposed artifact correction relies on the estimation of the 4M parameters used in the model of $\psi_m$. This estimation depends on the minimization of a multivariable function *C*, utilizing the interior point method for the optimization process (MATLAB's fmincon function).

This multivariable function *C* is defined from the model spectrum of the acquisition $\widehat{S(f)} = FT\left(\widehat{s_{acq}}\right)$ and the acquired spectrum: $S(f) = FT\left(s_{acq}\right)$, as well as the residual spectrum $\mathrm{K}(f) = S(f) - \widehat{S(f)}$.

The cost function *C*, which is used in the minimization $\min\limits_{A_m, \alpha_m, f_m, \phi_m} C(A_m, \alpha_m, f_m, \phi_m)$ is defined with two weighted terms:

$$C = C_1 + \beta C_2 + \gamma C_3 \quad [3]$$

$$C_1 = \sum_{f \epsilon [f_m]_{m=0 \ldots M}} \|K(f)\|_2^2 \quad [4]$$

with $[f_m]$ the frequency range around artifact frequencies $[f_m]$ =$[\frac{1}{m\Delta t} - 0.16\text{ppm} - \frac{1}{m\Delta t} + 0.16\text{ppm}]$ and

$$C_2 = \sum_{f \epsilon [f_i]} \left\| \frac{\partial^2 K(f)}{\partial f^2} \right\|_2^2 \text{ [5] and } C_3 = \frac{\max_{f \in [f_i]} |\widehat{S(f)}|}{\max_{f \in [f_i]} |S(f)|} \text{ [6]}$$

with $f_i$ the frequency range around artifact frequencies that fall into a spectral region of interest (Figure 2C). $C_1$ corresponds to the main objective of minimization, i.e., to model the artifactual signal and explore the possibility to reduce or eliminate it. This cost is concentrated on the frequency regions affected by the artifact. However, we have found that when artifact peaks are present in a frequency region where a component/metabolite of interest is expressed, we need to refine this cost function and further constrain the optimization. So, a term $C_2$ was added to account for the evolution of the K(f) amplitude variation rate.

In fact, this term prevents the large, rapid signal variations in K(f) that characterize artifacts compared with the smoother behavior of metabolite/biochemical compound peaks. The $C_2$ concentrates the calculation only on the frequency region where the metabolite signal needs to be preserved. Finally, a last term $C_3$ was found to be useful in practice - as many sets of parameters $A_m, \alpha_m, f_m, \phi_m$ are possible - to help the algorithm to converge toward realistic solutions. This last term constrains the maximum amplitude of the artifact as a ratio to the maximum amplitude of the signal in the frequency range *f*.

An iteration was performed according to the final value of the global cost function C. This was compared with a cost function $C_{noise}$ like the $C_1$ calculation, but for a frequency zone containing only noise. The minimization calculation was repeated as long as the global cost function is greater than $C_{noise}$, taking a closer solution as a starting point at each iteration. To optimize computational time and solution efficiency, a limit of 10 iterations was set. Retrospective correction was applied to the gradient-free acquisition case to assess its relevance regarding artifact behavior (Figure S3-S4 in supplementary material).

*In vitro and in vivo evaluation*

The retrospective artifact correction method was applied to spectra obtained on phantoms and spectra acquired in the lower leg of healthy volunteers (see table S1 for details of the acquisition parameters).

Spectra from each center of the object of interest (either phantom or calf) were used, and mean values ± standard deviation (SD) were reported for the SNR of water, lipids, and artifacts, the amplitude ratio between peaks, and the artifact reduction factor, as detailed next.

After applying this method, artifact reduction was characterized by a ratio derived from integrating the absolute value of the real part of the spectrum over a frequency range that encompasses the artifacts before and after correction. For artifacts present in a frequency region of interest, the impact of the correction on the quantification of the spectra was evaluated. To this end, the spectra were quantified using the LCModel (version 6.3-1R) quantification method (using the "muscle-5" prior knowledge basis set) before and after correction.

## Results

Figure 2 illustrates the variation of characteristic artifact resonance frequencies according to the number of temporal interleaves. In the case of temporal interleaving M = [4,5,7], the experimental result is a $\Delta f$ = [500,400,286] Hz (Fig. 2A-C), which is in line with the characterization of these artifacts detailed in Method (Origin of the artifact and proposed mathematical model).

In the time domain, the $\Delta T$ difference between signal jumps is also related to temporal interleaving (Figure 2B), such as $\Delta T = M \Delta t$.

For gradient-off acquisitions, the procedure described in Figure 3 shows how to extract and characterize the artifact by leveraging the fact that each interleave can be compared point-by-point to the first, and that the signal difference between the interleaves can be extracted. The effect on the spectra is the total removal of artifacts at the characteristic frequency (Figure 3B). On the FID, it no longer presents signal jumps due to the incoherence in the amplitude of temporal interleaves. Note that this is not a correction procedure, but rather a method to study the phenomenon.

According to Figure 3D, each interleave differs from the first interleave by an additional sinusoidal signal, of low amplitude and that appears to decay slowly. Considering that the first interleave might also exhibit this additional deviation, we generalized this modeling to all interleaves. So the amplitude parameter *Am* has been constrained in the minimization. The artifact amplitude is related to the amplitude of the deterministic real signal. For acquisition with water suppression, the artifact amplitude is then very small (Figure S5 in supplementary material).

For the phantom acquisitions, the ratio of the average artifact peaks and the water peak amplitudes was about 0.6%, and the artifact reduction achieved by the retrospective correction method was about a factor of 4 (Table 1).

| | **Phantom** | | | |
|---|---|---|---|---|
| M Temporal interleaves | 5 | 5 with correction | 7 | 7 with correction |
| σ noise | 0.0029 (0.0001) | 0.0030 (0.0001) | 0.0031 (0.0002) | 0.0031 (0.0002) |
| Baseline | 0.24 (0.01) | 0.24 (0.01) | 0.20 (0.01) | 0.20 (0.01) |
| SNR water | 44506.21 (4852.25) | 43910.44 (5381.09) | 41080.98 (5160.39) | 41004.91 (6866.78) |
| SNR artifacts | 322.24 (20.59) | 104.80 (6.01) | 246.08 (10.32) | 100.45 (7.95) |
| Artefact/water amplitude ratio % | 0.73 (0.08) | 0.24 (0.02) | 0.61 (0.07) | 0.25 (0.02) |
| Reduction factor | | 4.35 (0.49) | | 4.19 (0.38) |
| | **In Vivo** | | | |
| M Temporal interleaves | 4 | 4 with correction | 5 | 5 with correction |
| σ noise | 0.0075 (0.0013) | 0.0075 (0.0013) | 0.0057 (0.0006) | 0.0057 (0.0006) |
| Baseline | 0.33 (0.02) | 0.31 (0.02) | 0.37 (0.02) | 0.37 (0.02) |
| SNR water | 3337.06 (592.49) | 3256.09 (596.68) | 4397.53 (476.54) | 4395.57 (512.77) |
| SNR lipid | 199.04 (35.14) | 197.84 (34.92) | 261.77 (23.29) | 265.56 (38.40) |
| SNR artifacts | 73.58 (13.13) | 50.76 (8.01) | 86.47 (4.94) | 68.60 (7.40) |
| Lipid/water amplitude ratio % | 5.98 (0.48) | 6.10 (0.51) | 5.99 (0.56) | 6.08 (0.90) |
| Artifact/water amplitude ratio % | 2.21 (0.06) | 1.57 (0.07) | 1.98 (0.15) | 1.56 (0.03) |
| Artifact/lipid amplitude ratio % | 37.09 (3.02) | 25.80 (1.79) | 33.15 (1.74) | 26.28 (4.05) |
| Reduction factor | | 1.91 (0.33) | | 1.75 (0.16) |

***Table 1: SNR and peak amplitude ratio of artifacts and water according to the different acquisitions in phantoms and in healthy volunteer with and without retrospective correction of the artifacts.*** *The reduction factor characterized the artifact reduction after applying this method and is describe in Material and Method.*

In vivo acquisitions with a temporal interleaving number M = 5 and $\Delta t = 0.5ms$ present artifacts at $\frac{1}{M\Delta t} = 400\ Hz$ on both sides of the water peak (Figure 4). The average ratio of artifact peaks and the water peak amplitudes in the normal calf muscle (peaks extracted from a 3x3 voxels ROI, i.e., 9 spectra) was about 2%, and the artifact reduction, except for the region of interest, was about a factor of 1.8 (Table 1). In this context, the weighting parameters $\beta$ and $\gamma$, presented in equation [3], were optimized and set to ½.

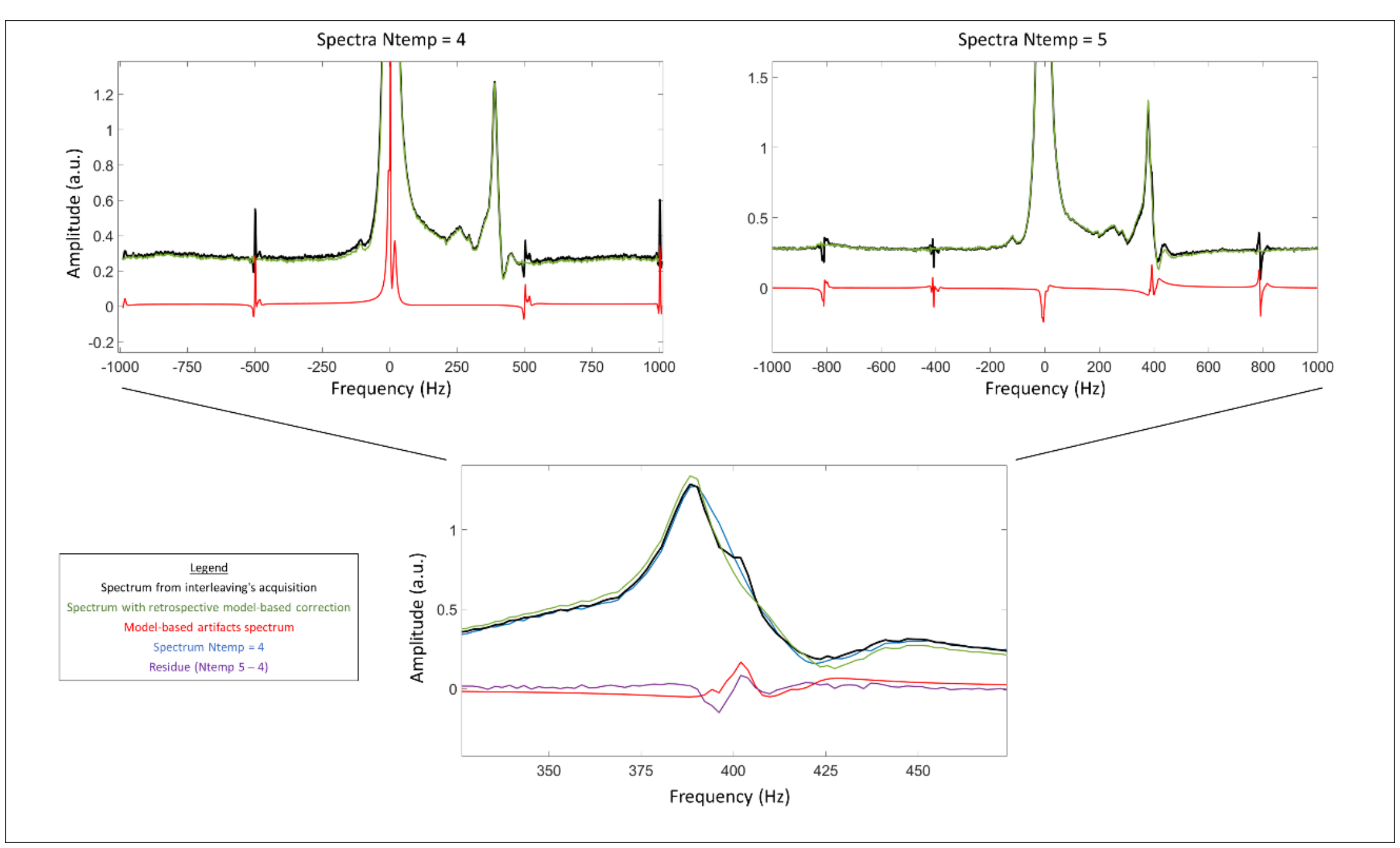

***Figure 4: Evaluation of the correction of artifacts in a region of interest (lipids) for 5 temporal interleaves in a spectrum obtained in the lower leg of a healthy volunteer.*** *Ground truth is obtained using an acquisition at exactly the same location and with 4 temporal interleaves, thus avoiding the presence of artifacts in the region of interest. In the lipid zone, artifacts are reduced by the correction.*

Since these artifacts occur within a lipid region of interest, the evaluation of our retrospective correction method was carried out by comparing the quantification of lipids present in this zone with and without correction to an acquisition with no artifacts in this lipid region (obtained by playing on the number of interleaves M = 4, $\frac{1}{M\Delta t} = 500Hz$). The correction particularly impacts the quantification of intra-myocellular lipids (IMCL). Indeed, the quantification results for spectra without artifacts in the lipid zone (M=4) did not find any IMCL contribution over the 9 processed spectra obtained from the manually selected 3x3 voxels ROI.

In contrast, LCModel's quantification of the spectra obtained with M=5 temporal interleaves (contaminated in the lipid spectral region) led to an erroneous detection of IMCL contribution in 7 cases over the 9 in-vivo spectra evaluated. After correction, the cleared dataset exhibits only 1 out of 9 spectra with minor incorrect small IMCL.

Figure 4 presents modeling and artifact correction with our method on acquisitions with and without artifacts in the lipid zone. The impact of the correction in the lipid zone is clearly visible, and the artifact's attenuation can be appreciated visually. Therefore, the results after correction offer 89% of the quantification result obtained on artifact-free spectra in the zone of interest. Acquisitions repeated on two different 3T scanners confirmed similar behavior, excluding site-specific hardware issues.

## Discussion and conclusion

All sequential tests presented with different modes and settings of spiral MRSI sequences or conventional FID sequences, demonstrated artifacts, as soon as temporal interlacing is used, whatever the configurations with temporal interleaving (MRSI with spiral gradient, without gradients, FID with temporal shift, conventional FID with repetition, or even with another scanner see supplementary material S6). Their non-repetitive nature, with a random amplitude in the time domain and a number of temporal interleave specific periodicities in the frequency domain make them impossible to measure for internal system calibration and compensation. All our tests also attested that no acquisition strategy can succeed in removing or tackling the issues.

The origin of artifacts caused by temporal amplitude/phase glitches could partly be due to the impulse response of the band-pass filtering process used in the ADC[24]. The issue could be particularly pronounced in spiral out readouts due to a high signal at the start of the readout when the center of the k-space is sampled. This contrasts with 1D readouts, such as echo planar imaging (EPI) or radial scanning, where the signal maximum is an echo at the center of the ADC. The impulse response, which is probably influenced by the input signal, could explain the spurious variation when the ADC sampling window is shifted in time. These errors are well known in electronics when attempting to increase bandwidth by temporally interleaving ADCs[14–16]. Also, both the RF transmitter and receiver chains require highly stable clock signals, as defined by strict jitter specifications. However, such information is often difficult to obtain or requires specific characterization[25]. Based on the experiments conducted in this paper, we hypothesize that minor glitches in the transmission or receive chains—potentially compounded by issues during the analog-to-digital conversion (ADC)—may contribute to the small deviations observed. Although subtle, these deviations would be sufficient to introduce artifacts in interleaved scans.

This issue has not been extensively discussed in the literature, as the artifacts are proportional to the amplitude of the sampled signal and become problematic when the number of interleaves M exceeds two. For M=2, they are confined to the edges of the spectral band, at the Nyquist frequency. In the case of acquiring a spectroscopic signal without water suppression, the amplitude of these fluctuations/random deviations is comparable to the spectral content other than water, such as lipids. Since most spectroscopic methods have been developed with water suppression to focus on metabolite signals, this issue has not received much attention as in such cases, the artifacts are often masked by noise. However, it is crucial to consider this challenge, as applications of spectroscopy without water suppression are of significant interest. Modeling and retrospective artifact suppression demonstrated excellent efficiency in phantom experiments, with a very high SNR, and during *in vivo* experiments outside the lipid region. When the artifacts occur in the lipid region, their influence on quantification results can be significant, and the correction method, even if not flawless in eliminating artifacts, proves valuable in minimizing quantification error. The model shown in Figure 4 presents a pattern that approximates the artifacts in the lipids without reliably suppressing them but prevents the appearance of a residual peak.

The decaying component of the artifacts appears to be linked to the fact that we are sampling a decaying signal and the jumps are related with the amplitude of the deterministic signal being sampled. The acquisitions were realized with a high signal-to-noise ratio, and according to the results, reporting the amplitude ratio of artifacts to water, acquisitions with a lower SNR may result in the disappearance of artifacts embedded in the noise. This assumption could be true for acquisitions made with X-nuclei, such as 31P. Note that suppressing the water peak would not influence the occurrences of the artifacts, as they stem from incoherence between the temporal interleaves. The artifacts are now well-defined and can be positioned to avoid interference with frequency regions of interest (e.g., in Figure 4 with M=4 and M=5). However, constraints related to the system, such as slew rate, and constraints related to the desired investigation (FOV, spatial and temporal resolution, acquisition time) can limit the ability to design the acquisition protocol. Within this framework, our approach provides a method to attenuate artifact impairment in frequency regions of interest when hardware or other constraints limit the choice of acquisition protocol.

In conclusion, we suggest leveraging the artifacts' periodic and limited–frequency nature. This led us to propose either i) to optimize the acquisition protocol by choosing the appropriate number of temporal interleaves to prevent contamination within the frequency range of interest or ii) to design a retrospective correction method that minimizes a cost function to model and compute a first-order approximation of the artifacts within the limited frequency range where they occur.

**Acknowledgments**

The authors thank Josef Pfeuffer, Application Development, Siemens Healthineers AG, Erlangen for providing the spiral framework source code. They also express their gratitude to Hervé Saint Jalmes for his insightful discussions on the excitation/detection chain of an MRI scanner. This work was partly supported by the LABEX PRIMES (ANR-11-LABX-0063) and Siemens Healthineers. It was also partly performed on a facility of France Life Imaging network (grant ANR-11-INBS-0006 / ROR identifier: https://ror.org/01vzmdx15)

**Author Contributions:**

**Antoine Naegel:** conceptualization, data curation, formal analysis, methodology, investigation, software, visualization, validation, writing – original draft, writing – review and editing.

**Magalie Viallon:** investigation, methodology, funding acquisition , project administration , supervision, validation, writing – review and editing.

**Thomas Troalen**: investigation, software, review and editing.

**Christopher T. Rodgers**: investigation,writing, review and editing.

**Dinesh K Deelchand**: investigation, writing, review and editing.

**Paul Nobre:** investigation, validation, review and editing.

**David Porter:** investigation, writing, review and editing.

**Jabrane Karkouri,** formal analysis, software, visualization, validation, review and editing.

**Hélène Ratiney:** conceptualization, investigation, methodology, funding acquisition, project administration , software, supervision, validation, original draft, writing – review and editing.

## Data Availability

The experimental data can be provided upon request from the authors.

## Conflict of Interest

The authors declare that they have no conflict of interest.

**Tables :**

| | Phantom | | | |
|---|---|---|---|---|
| M Temporal interleaves | 5 | 5 with correction | 7 | 7 with correction |
| σ noise | 0.0029 (0.0001) | 0.0030 (0.0001) | 0.0031 (0.0002) | 0.0031 (0.0002) |
| Baseline | 0.24 (0.01) | 0.24 (0.01) | 0.20 (0.01) | 0.20 (0.01) |
| SNR water | 44506.21 (4852.25) | 43910.44 (5381.09) | 41080.98 (5160.39) | 41004.91 (6866.78) |
| SNR artifacts | 322.24 (20.59) | 104.80 (6.01) | 246.08 (10.32) | 100.45 (7.95) |
| Artefact/water amplitude ratio % | 0.73 (0.08) | 0.24 (0.02) | 0.61 (0.07) | 0.25 (0.02) |
| Reduction factor | | 4.35 (0.49) | | 4.19 (0.38) |
| | **In Vivo** | | | |
| M Temporal interleaves | 4 | 4 with correction | 5 | 5 with correction |
| σ noise | 0.0075 (0.0013) | 0.0075 (0.0013) | 0.0057 (0.0006) | 0.0057 (0.0006) |
| Baseline | 0.33 (0.02) | 0.31 (0.02) | 0.37 (0.02) | 0.37 (0.02) |
| SNR water | 3337.06 (592.49) | 3256.09 (596.68) | 4397.53 (476.54) | 4395.57 (512.77) |
| SNR lipid | 199.04 (35.14) | 197.84 (34.92) | 261.77 (23.29) | 265.56 (38.40) |
| SNR artifacts | 73.58 (13.13) | 50.76 (8.01) | 86.47 (4.94) | 68.60 (7.40) |
| Lipid/water amplitude ratio % | 5.98 (0.48) | 6.10 (0.51) | 5.99 (0.56) | 6.08 (0.90) |
| Artifact/water amplitude ratio % | 2.21 (0.06) | 1.57 (0.07) | 1.98 (0.15) | 1.56 (0.03) |
| Artifact/lipid amplitude ratio % | 37.09 (3.02) | 25.80 (1.79) | 33.15 (1.74) | 26.28 (4.05) |
| Reduction factor | | 1.91 (0.33) | | 1.75 (0.16) |

***Table 1: SNR and peak amplitude ratio of artifacts and water according to the different acquisitions in phantoms and in healthy volunteer with and without retrospective correction of the artifacts.*** *The reduction factor characterized the artifact reduction after applying this method and is describe in Material and Method.*

# Supplementary material:

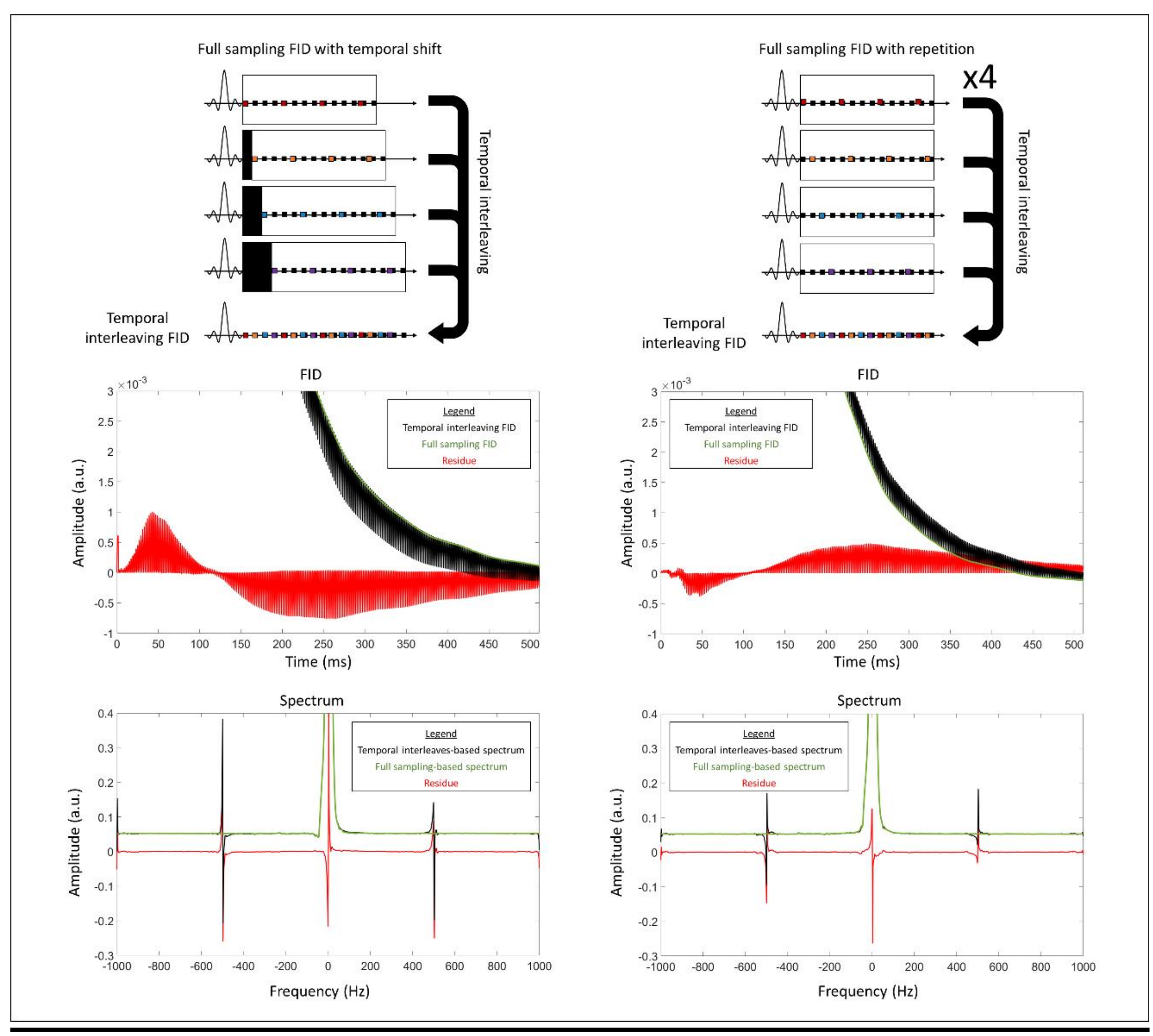

***Figure S1: Conventional FID spectroscopy sequence with time shifting (left) or repetition (right) to create temporal interleaving.*** *Artifacts are created a posteriori by time interleaving to reconstruct the temporal dimension and can be compared with the acquisition of a full sample with no artifacts. For each type of acquisition, 4 dummy scans are performed beforehand, and the repeated acquisition is also averaged 4 times per interleaving (4 cycles of 4 acquisitions to reconstruct 4 interleaved signals), producing an acquisition of 4x4 acquisitions.*

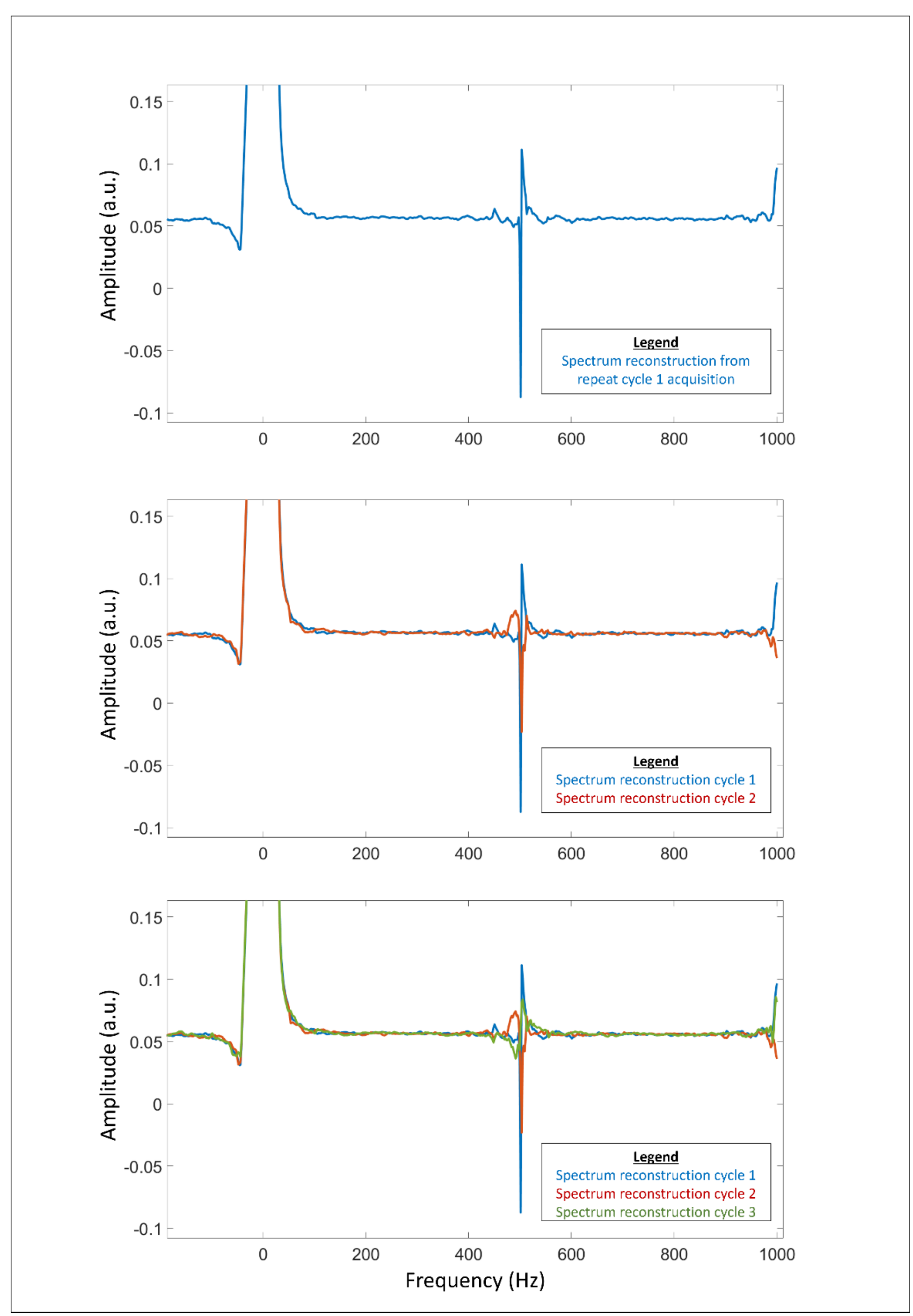

***Figure S2: Illustration of the random nature of the artifacts.*** *The repetitive sequence described in figure S1 is used to reconstruct time-interleaved data every 4 repetitions over 5 cycles (20 acquisitions in total). The first 3 cycles are shown here, superimposed from top to bottom. The artifacts show the same frequencies but not the same behavior.*

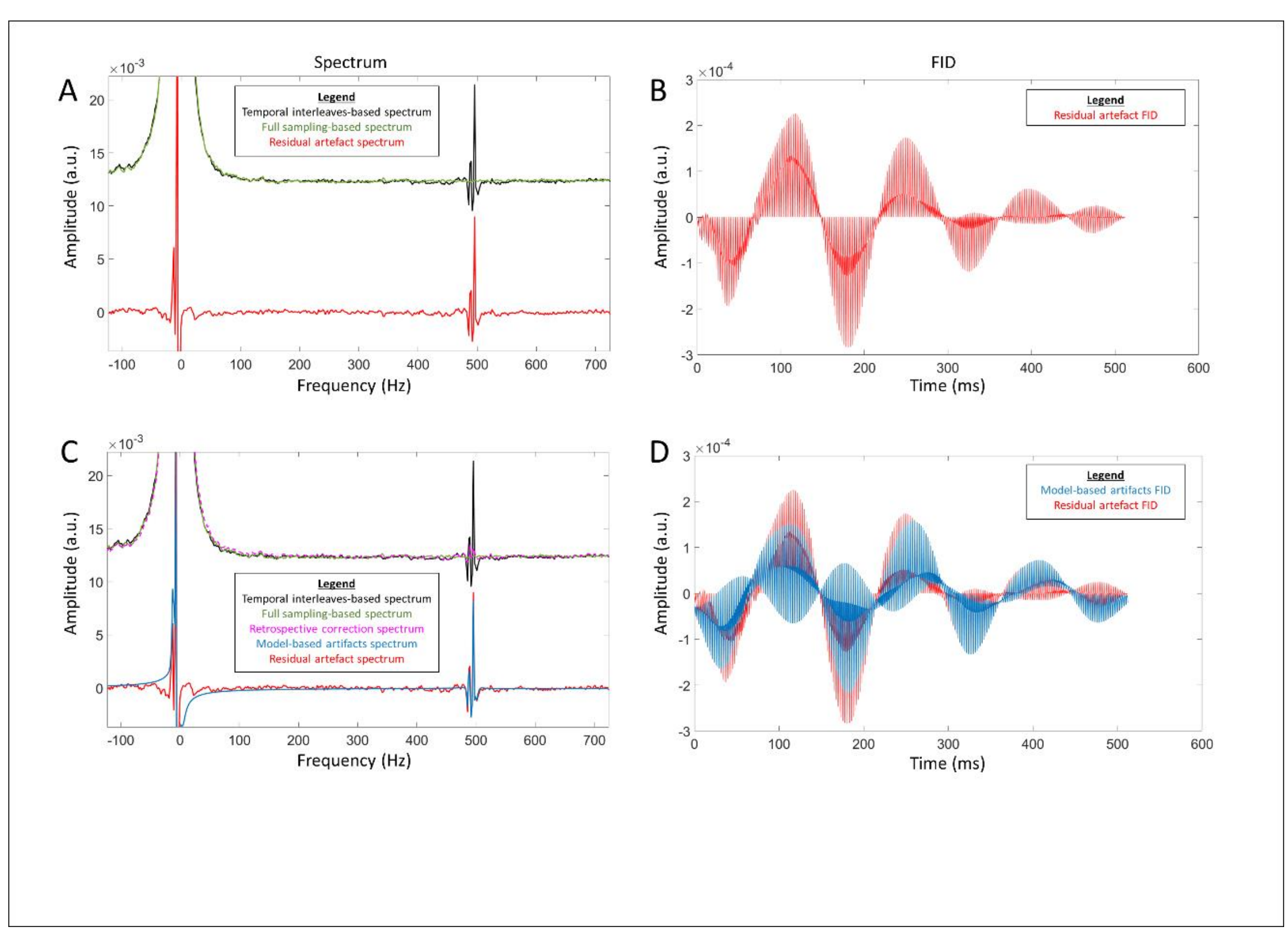

***Figure S3: Application of the retrospective artifact correction method to the gradient-free acquisition used to characterize artifacts (see figure 3).*** *A-B, application of the ratio correction method to the gradient-free FID acquisition. C-D, application of the retrospective correction method to the gradient-free FID acquisition, illustrating the model-based correction method's ability to capture artifact behavior.*

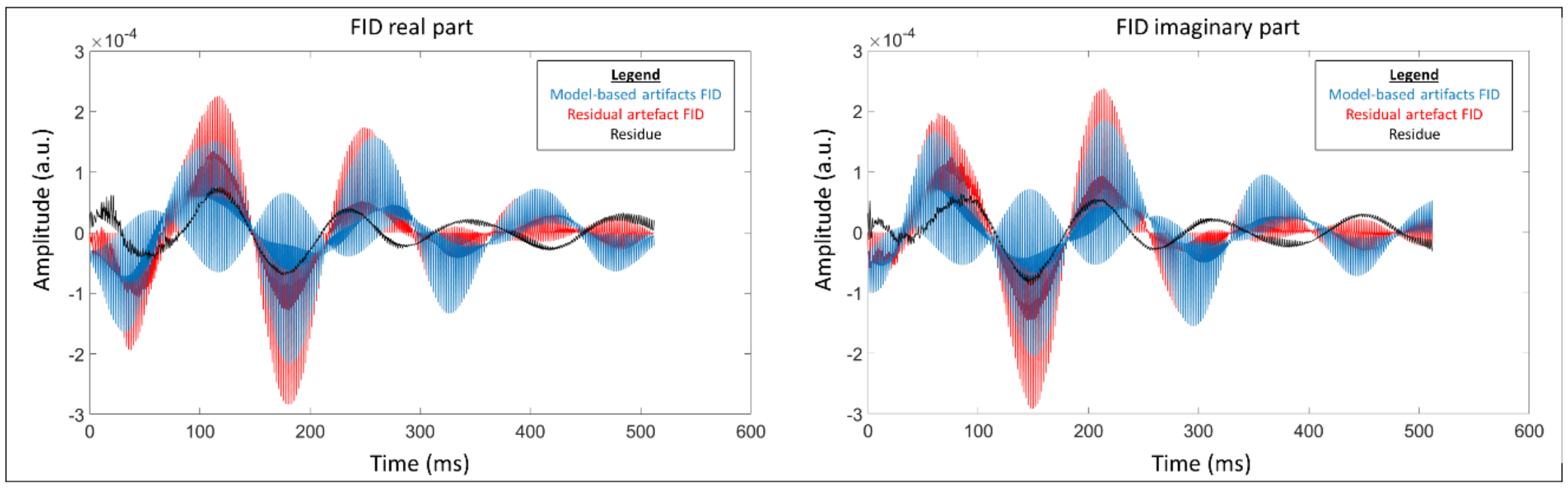

***Figure S4: The model-based artifacts FID, the residual artifact FID obtained via the first full sampling interleave and related residue.*** *Left, the real part and, Right, the imaginary part*

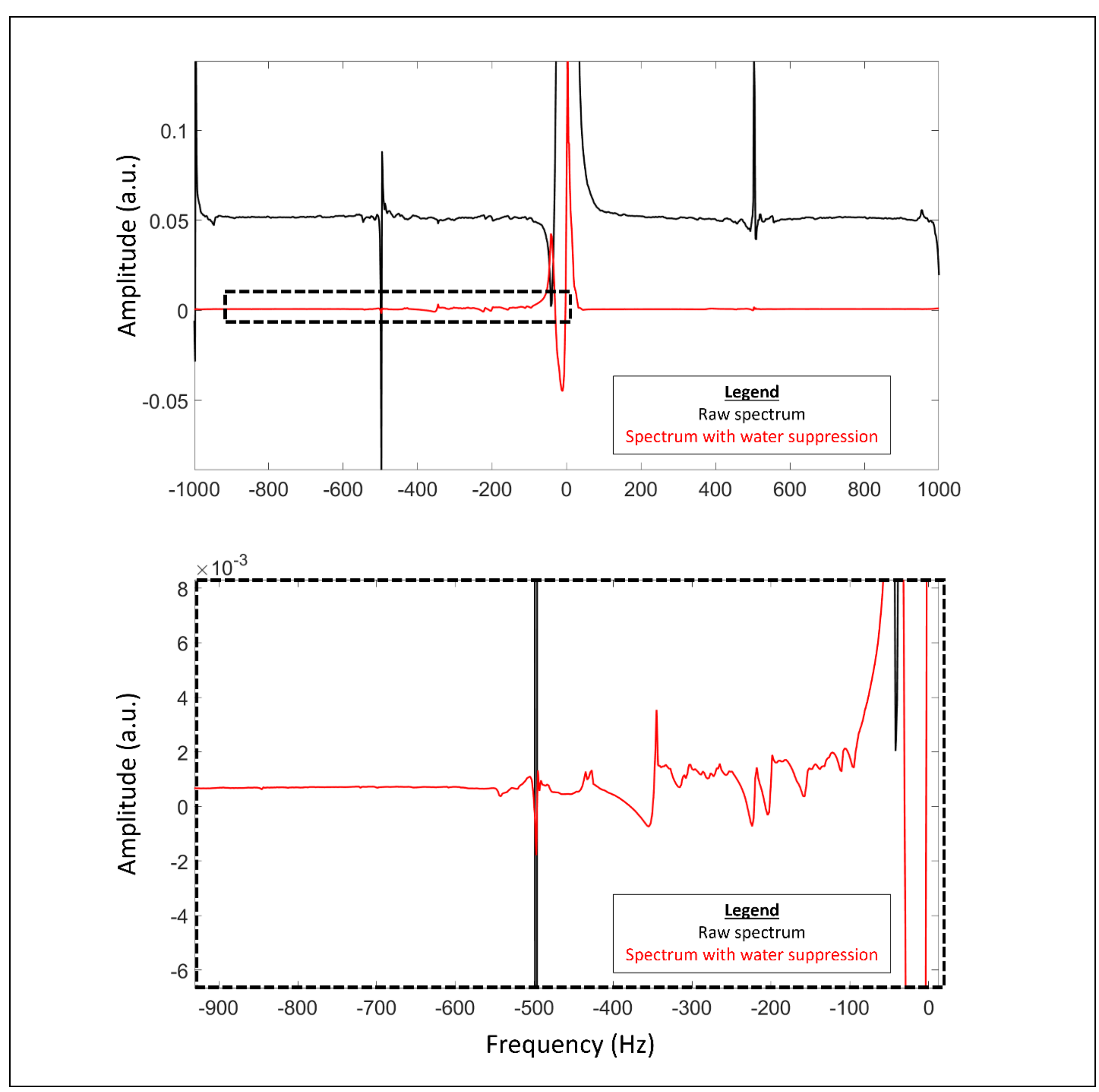

***Figure S5: Influence of water suppression on temporal interleave artifacts.*** *Conventional FID acquisition with temporal shift and reconstruction of the spectroscopic dimension with or without water suppression. Artifacts are reduced in amplitude but still present at the frequency identified for a temporal interleave number of 4 (see bottom graph for zoom on spectrum with water suppression).*

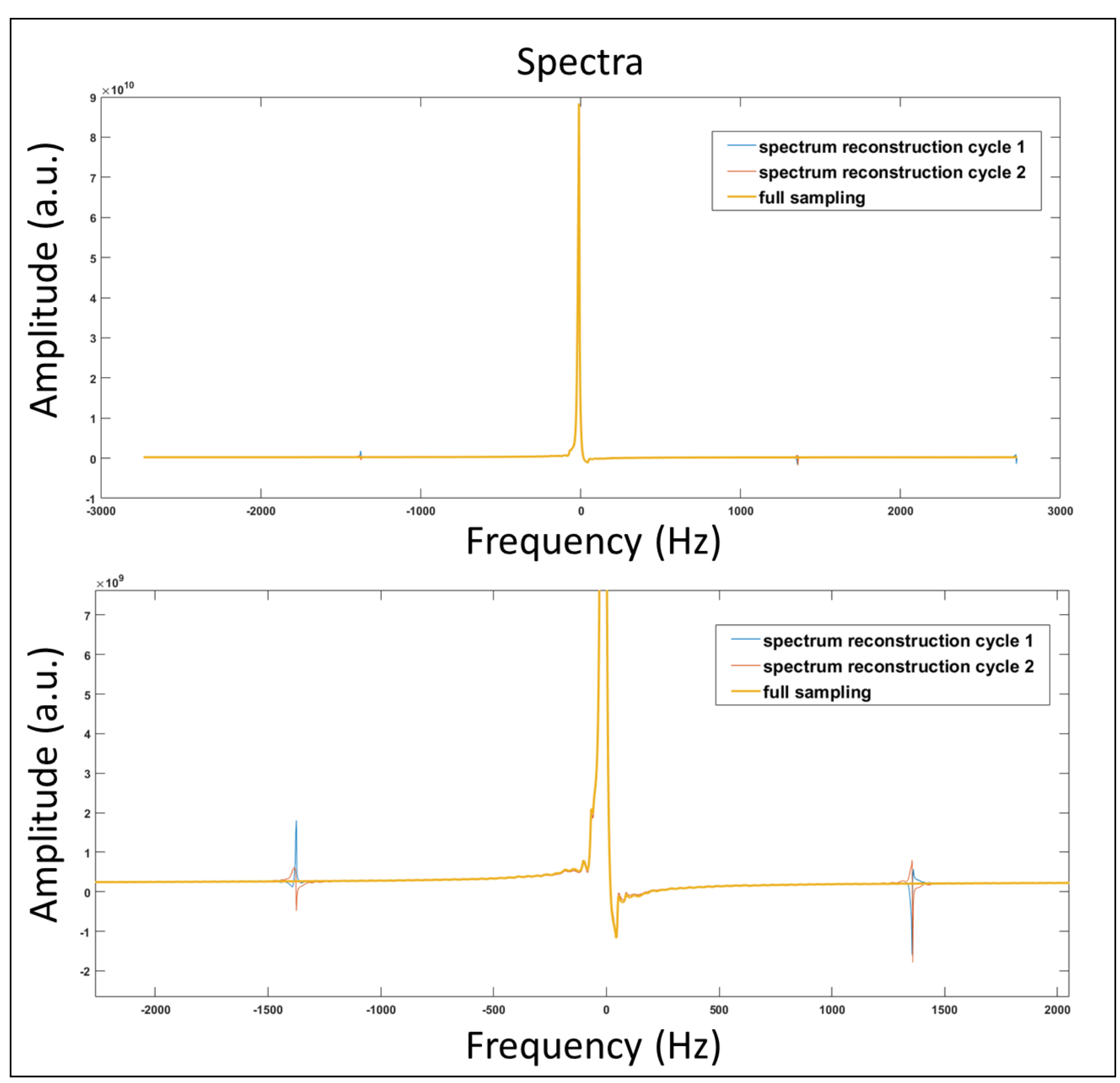

***Figure S6: PRESS data acquired without water suppression on a phantom containing distilled water at 11.7 T Bruker MRI scanner****, band with 64 accumulations, each accumulation is then divided into packets of 16, to simulate 4 interleaving and reconstruct a temporal interleaving-based signal. ),Cycle 1 and cycle 2 are two different ways of arranging 16x4 spectra, demonstrating the random nature of artifacts. The figure in the bottom is a zoom on the artifact.*

| | **Phantom** | **In vivo** |
|---|---|---|
| TR/TE (ms) | 5000/2 | 2000/2 |
| Flip angle (°) | 90 | 90 |
| Dummy scan | 4 | 4 |
| FOV ($mm^3$) | 250x250x10 | 200x200x10 |
| Matrix size | 32x32 | 64x64 |
| Voxel size ($mm^3$) | 7.8x7.8x10 | 3.1x3.1x10 |
| $\Delta t$ Temporal resolution (us) | 500 | 500 |
| $\frac{1}{\Delta t}$ Spectral bandwidth | 2000 | 2000 |
| $M$ Temporal interleaves | 7 / 5 | 4 / 5 |
| Spatial interleaves | 4 / 7 | 26 |
| Slewrate (mT/(m*ms)) | 80 | 120 / 100 |
| ADC length (us) | 3040 / 2080 | 1920 / 2240 |
| Spiral length (us) | 2820 / 1920 | 1820 / 2000 |
| $\Delta T$ Interleaves resolution (us) | 3500 / 2500 | 2000 / 2500 |
| Frequency delta between artefactual peaks (Hz) | 286 / 400 | 400 / 500 |
| Total Time Acquisition | 2’22’’ / 2’55’’ | 3’28’’ / 4’20’’ |

***S1: Summary of the main sequence acquisition parameters performed on phantom and in vivo.***